\documentclass[10pt]{article}
\usepackage{amsfonts}
\usepackage{amsfonts}
\usepackage{}
\usepackage{amsfonts}
\usepackage{amsmath}
\usepackage{amssymb, amscd, amsthm}
\usepackage[all]{xy}
\usepackage[dvips]{graphicx}
\usepackage{verbatim}
\usepackage{cite}
\usepackage{enumerate}
\usepackage{float}

\newtheorem{theorem}{Theorem}
\theoremstyle{plain}

\newtheorem{example}[theorem]{Example}
\newtheorem{lemma}[theorem]{Lemma}

\numberwithin{equation}{section}
\usepackage{longtable}

\usepackage{multirow}

\begin{document}

\baselineskip 17pt\title{\Large\bf Construction on Griesmer Codes with Dimension Not Less than Five}
\author{Wen Ma\footnote{mawen95@126.com (W.~Ma)} \qquad Jinquan Luo \footnote{Corresponding author,luojinquan@mail.ccnu.edu.cn (J.~Luo).
\newline
 The authors are with School of Mathematics and Statistics $\&$ Hubei Key Laboratory of Mathematical Sciences, Central China Normal University, Wuhan, China 430079.}}
\date{}
\maketitle

{\bf Abstract}: In this paper we construct new Griesmer codes of dimension $k\geq 5$ by means of some geometric methods such as projective dual and geometric puncturing.

{\bf Key words}: Projective geometry; Divisible codes; Linear codes; Projective dual
\section{Introduction}

 \quad\; We denote by $\mathbb{F}_q^n$ the vector space of $n$-tuples over $\mathbb{F}_q$, the field of $q$ elements. The \emph{weight} of a vector $\textbf{c}\in \mathbb{F}_q^n$, denoted by $\omega t(\textbf{c})$, is the number of nonzero coordinate positions in $\textbf{c}$. An $[n,k,d]_q$ code $\mathcal {C}$ is a $\mathbb{F}_q$-linear subspace of $\mathbb{F}_q^n$ with dimension $k$ and  minimum
 (Hamming) distance $d$.

 A fundamental problem in coding theory is to  determine $n_q(k,d)$, the minimum value of $n$ for which there exists an $[n,k,d]_q$ code for given $q,k$ and $d$, see [\ref{RHO},\ref{RHE}]. The values of $n_q(k,d)$ are determined for all $d$ only for some small values of $q$ and $k$ [\ref{MG},\ref{TMGBFL}]. As a lower bound on $n_q(k,d)$, the Griesmer bound (see [\ref{JH}]) states that:
 $$n\geq g_q(k,d)=\sum\limits_{i=0}^{k-1}\lceil d/q^i\rceil,$$
 where $\lceil x\rceil$ is the ceiling function. The code $\mathcal {C}$ is called \emph{Griesmer} if it attains the Griesmer bound.

 The main aim of this paper is to construct new Griesmer codes of dimension $k\geqslant 5$, which are depicted as follows.

\begin{theorem}\label{the1}
There exists $[g_q(k,d),k,d]_q$ code for $(k-3)q^{k-1}-3q^{k-2}+q^{k-3}-q^2+q\leq d\leq (k-3)q^{k-1}-3q^{k-2}+q^{k-3}$ with $k\geq 5$ and $q\geq k-2$.
\end{theorem}

\begin{theorem}\label{the2}
There exists $[g_q(k,d),k,d]_q$ code for $(k-2)q^{k-1}-5q^{k-2}+2q^{k-3}-q^2+q\leq d\leq (k-2)q^{k-1}-5q^{k-2}+2q^{k-3}$ with $k\geq 5$ and $q\geq \max\{5,k-2\}$.
\end{theorem}

\section{Preliminaries}

\quad We denote by ${\rm PG}(r,q)$ the projective space of dimension $r$ over $\mathbb{F}_q$ . A $j$-flat is a projective subspace of dimension $j$ in ${\rm PG}(r,q)$. The 0-flats, 1-flats and $(r-1)$-flats in ${\rm PG}(r,q)$ are called points, lines, and hyperplanes respectively. We denote by $\theta_j$ the number of points in a $j$-flat, i.e., $\theta_j=\frac{q^{j+1}-1}{q-1}$.

Let $\mathcal {C}$ be an $[n,k,d]_q$ code with full support, i.e., there is no coordinate in which all the codewords have entries 0. The columns of a generator matrix of $\mathcal {C}$ can be considered as a multiset of $n$ points in $\Sigma={\rm PG}(k-1,q)$, denoted by $\mathcal {M}_{\mathcal {C}}$. An $i$-point is a point of $\Sigma$ which has multiplicity $i$ in $\mathcal {M}_{\mathcal {C}}$. Denote by $\gamma_0$ the maximum multiplicity of a point from $\Sigma$ in $\mathcal {M}_{\mathcal {C}}$. Let $C_i$ be the set of $i$-points in $\Sigma$, $0\leq i\leq \gamma_0$. Let $\lambda_i=|C_i|$, where $|C_i|$ denotes the number of elements in the set $C_i$. For any subset $S$ of $\Sigma$, the multiplicity of $S$ with respect to $\mathcal {M}_{\mathcal {C}}$, denoted by $m_{\mathcal {C}}(S)$, is defined as $m_{\mathcal {C}}(S)=\sum_{i=0}^{\gamma_0}i\cdot |S\cap C_i|$. Then we obtain the partition $\Sigma=\bigcup_{i=0}^{\gamma_0} C_i$ such that $n=m_{\mathcal {C}}(\Sigma)$ and
$$n-d=max\{m_{\mathcal {C}}(\pi)\mid \pi\in \mathcal {F}_{k-2}\},$$
where $\mathcal {F}_j$ denotes the set of $j$-flats of $\Sigma$. Conversely, such a partition $\Sigma=\bigcup_{i=0}^{\gamma_0} C_i$ as above gives an $[n,k,d]_q$ code in a natural manner.  Denote by $a_i$ the number of hyperplanes with multiplicity $i$ in $\Sigma$. The list of the values $a_i$ is called the spectrum of $\mathcal {C}$. An $[n,k,d]_q$ code is called $m$-divisible if all codewords have weights divisible by  integer $m>1$.

\begin{lemma}([\ref{MTKO}]).
Let $\mathcal {C}$ be an $m$-divisible $[n,k,d]_q$ code with $q=p^h$, $p$ prime, whose
spectrum is
$$(a_{n-d-(\omega-1)m},a_{n-d-(\omega-2)m},\cdots,a_{n-d-m},a_{n-d})=(\alpha_{\omega-1},
\alpha_{\omega-2},\cdots,\alpha_1,\alpha_0),$$
where $m=p^r$ for some $1\leq r\leq h(k-2)$ satisfying $\lambda_0>0$ and
$$\bigcap_{H\in\mathcal {F}_{k-2},m_\mathcal {C}(H)<n-d}H=\emptyset.$$
Then there exists a $t$-$divisible$ $[n^{*},k,d^{*}]_q$ code $\mathcal {C}^{*}$ with $t=q^{k-2}/m$, $n^{*}=\sum_{j=0}^{\omega-1}j\alpha_j=ntq-\frac{d}{m}\theta_{k-1}$, $d^{*}=((n-d)q-n)t$ whose spectrum is
$$(a_{n^{*}-d^{*}-\gamma_0t},a_{n^{*}-d^{*}-(\gamma_0-1)t},\cdots,a_{n^{*}-d^{*}-t},
a_{n^{*}-d^{*}})=(\lambda_{\gamma_0},\lambda_{\gamma_0-1},\cdots,\lambda_1,\lambda_0).$$
\end{lemma}

Note that a generator matrix for $\mathcal {C}^{*}$ is given by considering $(n-d-jm)$-hyperplanes as $j$-points in the dual space $\Sigma^{*}$ of $\Sigma$ for $0\leq j\leq\omega-1$, see [\ref{MTKO}]. The code $\mathcal {C}^{*}$ is called a \emph{projective dual} of $\mathcal {C}$, see [\ref{RHE}] and [\ref{AEB}].

\begin{lemma}([\ref{TMCO}],[\ref{TMYO}])\label{t=1}.
Let $\mathcal {C}$ be an $[n,k,d]_q$ code and let $\bigcup_{i=0}^{\gamma_0} C_i$ be the partition of $\Sigma={\rm PG}(k-1,q)$ obtained from $\mathcal {C}$.
If $\bigcup_{i\geq 1} C_i$ contains a $t$-flat $\Delta$ and if $d>q^t$, then there
exists an $[n-\theta_t,k,d^{'}]_q$ code $\mathcal {C}^{'}$ with $d^{'}\geq d-q^t$.
\end{lemma}

The code $\mathcal {C}^{'}$ in Lemma \ref{t=1} can be constructed from $\mathcal {C}$ by removing the $t$-flat $\Delta$ from the multiset $\mathcal {M}_{\mathcal {C}}$.
In general, the method for constructing new codes from a given $[n,k,d]_q$ code by deleting the coordinates corresponding to some geometric object in ${\rm PG}(k-1,q)$ is called \emph{geometric puncturing} [\ref{TMCO}].

The following results are known for $k=5$.

\begin{theorem}([\ref{TMK5}])\label{k51}.
There exists $[g_q(5,d),5,d]_q$ code for $2q^4-3q^3+q\leq d\leq 2q^4-3q^3+q^2$ and $q\geq 3$.
\end{theorem}

\begin{theorem}([\ref{TMK5}])\label{k52}.
There exists $[g_q(5,d),5,d]_q$ code for $3q^4-5q^3+q^2+q\leq d\leq 3q^4-5q^3+2q^2$ and $q\geq 5$.
\end{theorem}
In this paper, we generalize these two results to the case $k\geqslant 5$, see Theorems 1 and 2.
\section{Proof of Theorems 1 and 2}

\quad A set $S$ of $s$ points in ${\rm PG}(r,q)$, $r\geq2$, is called an $s$-$arc$ if no $r+1$ points are on the same hyperplane, see [\ref{JWPHP}] and [\ref{JWPHJ}] for more knowledge on arcs. One can take a normal rational curve as a $(q+1)$-arc when $q\geq r$, see [\ref{JWPHJ}].
We first assume $k\geq 4$ and $q\geq k-2$. Let $H$ be a hyperplane of $\Sigma={\rm PG}(k-1,q)$.
Take a $(q+1)$-arc $K=\{P_0,P_1,\cdots,P_q\}$ in $H$ and a line $l_0=\{P_0,Q_1,\cdots,Q_q\}$ of $\Sigma$ not contained in $H$ and meeting $H$ at the point $P_0$. Let $l_i$ be the line joining $P_i$ and $Q_i$ for $1\leq i\leq q$. Setting $C_1=(\cup_{i=1}^{q}l_i)\setminus l_0$, $C_{q-1}=\{P_0\}$, $C_0=\Sigma\setminus(C_1\cup C_{q-1})$, we get the following.

\begin{lemma}([\ref{TMK5}])
For $k\geq 4$, $q\geq k-2$, a $q$-divisible $[q^2+q-1,k,q^2-(k-3)q]_q$ code exists.
\end{lemma}

\begin{lemma}
There exists a $q$-divisible $[q^2+2q-1,k,q^2-(k-4)q]_q$ code $\mathcal {C}_1$ with
\begin{equation*}
a_{(k-2)q-1}=\binom{q}{k-4}+\binom{q}{k-3}.
\end{equation*}

\end{lemma}

\begin{proof}
For any $k\geq 6$ and a primitive element $\alpha\in\mathbb{F}_q^{*}$, take a normal rational curve $K=\{P_0,P_1,\cdots,P_q\} $ with
\begin{center}
$P_0(1,0,\cdots,0,0,0),\ P_i(1,\alpha^i,\alpha^{2i},\cdots,\alpha^{(k-3)i},\alpha^{(k-2)i},0)$ for $1\leq i\leq q-1$, $P_q(0,0,\cdots,0,1,0)$
\end{center}
 and the line $l_0=\{P_0, Q_1,\cdots,Q_q\}$ with
\begin{center}
$Q_i(1,0,\cdots,0,\alpha^{i})$ for $1\leq i\leq q-1$, $Q_q(0,0,\cdots,0,1).$
\end{center}
Note that $K\subseteq H=[0,0,\cdots,0,1]$, where $[a_0,a_1,\cdots,a_{k-1}]$ stands for the hyperplane in ${\rm PG}(k-1,q)$ defined by the equation $a_0x_0+a_1x_1+\cdots+a_{k-1}x_{k-1}=0$.
Let $H_{i_1i_2\cdots i_{k-3}}$ be the $(k-2)$-flat containing $l_{i_1}$, $l_{i_2},\cdots,l_{i_{k-3}}$ for $1\leq i_1< i_2< \cdots< i_{k-3}\leq q$. Take the point $Q(0,1,0,\cdots,0,1)$ and the plane $\delta_0=<l_0,Q>$, where $<\chi_1,\chi_2,\cdots>$ denotes the smallest flat containing $\chi_1,\chi_2,\cdots$.
For any point $P(a,b,0,\cdots,0,c)\in\delta_0$, the $(k-2)$-flat
\begin{eqnarray*}
  H_{i_1i_2\cdots i_{k-4}q}&=&\Big[0,
(-1)^{k-4}\alpha^{\sum\limits_{t=1}^{k-4}i_t},
(-1)^{k-5}\sum\limits_{1\leq s_1<s_2<\cdots <s_{k-5}\leq k-4}
\alpha^{i_{s_1}+\cdots+i_{s_{k-5}}},\\
      &&(-1)^{k-6}\sum\limits_{1\leq s_1< s_2< \cdots< s_{k-6}\leq k-4}
\alpha^{i_{s_1}+\cdots+i_{s_{k-6}}},
\cdots,(-1)\sum\limits_{t=1}^{k-4}\alpha^{i_t},
1,0,0\Big]
\end{eqnarray*}
contains $P$ if and only if $b=0$, i.e., $P\in l_0$ for $1\leq i_1< i_2<\cdots< i_{k-4}\leq q-1$. Similarly, the $(k-2)$-flat
$$H_{i_1i_2\cdots i_{k-3}}=\Big[0,(-1)^{k-3}\alpha^{\sum\limits_{t=1}^{k-3}i_t},
(-1)^{k-4}\sum\limits_{1\leq s_1<s_2< \cdots<s_{k-4}\leq k-3}
\alpha^{i_{s_1}+\cdots+i_{s_{k-4}}},\cdots,
(-1)\sum\limits_{t=1}^{k-3}\alpha^{i_t},1,0\Big]$$
contains $P$ if and only if $P\in l_0$ for $1\leq i_1< i_2< \cdots< i_{k-3}\leq q-1$.
Therefore, $H_{i_1i_2\cdots i_{k-3}}\cap \delta_0=l_0$ for $1\leq i_1< i_2< \cdots <i_{k-3}\leq q$ and no $(k-2)$-flat with multiplicity $((k-2)q-1)$ contains $Q$.
Hence we get a $q$-divisible $[q^2+2q-1,k,q^2-(k-4)q]_q$ code by adding $Q$ as a $q$-point, say $\mathcal{C}_1$.
The $(k-2)$-flats with multiplicity $((k-2)q-1)$ consist of
\begin{enumerate}[(i)]
\item the $\binom{q}{k-4}$ many $(k-2)$-flats $<l_{i_1},l_{i_2},\cdots,l_{i_{k-4}},Q>$, $1\leq i_1< i_2< \cdots< i_{k-4}\leq q$,
\item the $\binom{q}{k-3}$ many $(k-2)$-flats $H_{i_1i_2\cdots i_{k-3}}$, $1\leq i_1< i_2<\cdots<i_{k-3}\leq q$.
\end{enumerate}
Hence $a_{(k-2)q-1}=\binom{q}{k-4}+\binom{q}{k-3}$.
\end{proof}

Considering the $(k-2)$-flats with multiplicity $((k-2-j)q-1)$ in $\Sigma$ as $j$-points in $\Sigma^{*}$ for $j=0,1,2,\cdots,k-3$, we get the following $q^{k-3}$-divisible code $\mathcal {C}_1^{*}$ as a projective dual of $\mathcal {C}_1$.

\begin{lemma}
There exists a $q^{k-3}$-divisible $[2q^{k-1}-q^{k-2}+1+(k-5)\theta_{k-1},k,(k-3)q^{k-1}-3q^{k-2}+q^{k-3}]_q$ code $\mathcal {C}_1^{*}$.
\end{lemma}

\begin{lemma}\label{SK1}
The multiset $\mathcal {M}_{\mathcal {C}_1^{*}}$ contains $(q-1)$ disjoint lines.
\end{lemma}

\begin{proof}
Note that the points with multiplicity 0 for $\mathcal {C}_1^{*}$ are the $(k-2)$-flats with multiplicity $((k-2)q-1)$ for $\mathcal {C}_1$. Since $l_0^{*}$ is contained in $H_{i_1i_2\cdots i_{k-3}}$ and $<l_{i_1},l_{i_2},\cdots,l_{i_{k-4}},Q>$ in $\Sigma$, the $(k-3)$-flat $l_0^{*}$ contains exactly $\binom{q}{k-4}+\binom{q}{k-3}$ many points with multiplicity 0 in $\Sigma^{*}$.
Hence the number of points with multiplicity $i\geq 1$ on $l_0^{*}$ is $\theta_{k-3}-\binom{q}{k-4}-\binom{q}{k-3}\geq q-1$.
Recall that the $(k-3)$-flat $l_0^{*}$ is contained in the $(k-2)$-flats $P_0^{*}$ and $Q_1^{*},\cdots, Q_q^{*}$ in $\Sigma^{*}$. One can take $q-1$ skew lines in the $(k-2)$-flat $Q_1^{*}$ containing no point with multiplicity 0 in $\Sigma^*$.
\end{proof}

Setting $C_1=(\cup_{i=1}^{q-1}l_i)\setminus l_0$, $C_{q-1}=\{P_0,Q_q\}$, $C_{q}=\{P_q\}$, $C_0=\Sigma\setminus (C_1\cup C_{q-1}\cup C_q)$, we get the following.

\begin{lemma}([\ref{TMK5}])
For $k\geq 4$, $q\geq k-2$, a $q$-divisible $[q^2+2q-2,k,q^2-(k-3)q]_q$ code exists.
\end{lemma}

\begin{lemma}
There exists a $q$-divisible $[q^2+3q-2,k,q^2-(k-4)q]_q$ code $\mathcal {C}_2$ with
\begin{equation*}
a_{(k-1)q-2}=\binom{q-1}{k-3}+2\binom{q-1}{k-4}+\binom{q-1}{k-5}. \tag{b}
\end{equation*}

\end{lemma}

\begin{proof}
For any $k\geq 6$, take the $(q+1)$-arc $K$ and the line $l_0$ as for $\mathcal {C}_1$. Similarly to the situation for constructing $\mathcal {C}_1$, no $(k-2)$-flat with multiplicity $((k-1)q-2)$ contains $Q$.
Therefore, adding $Q$ as a $q$-point we get a $q$-divisible $[q^2+3q-2,k,q^2-(k-4)q]_q$ code, say $\mathcal {C}_2$. The $(k-2)$-flats with multiplicity $((k-1)q-2)$ contain
\begin{enumerate}[(i)]
\item the $\binom{q-1}{k-5}$ many $(k-2)$-flats
$<l_0,l_{i_1},l_{i_2},\cdots,l_{i_{k-5}},P_q,Q>$ with $1\leq i_1< i_2< \cdots< i_{k-5}\leq q-1$,
\item the $\binom{q-1}{k-4}$ many $(k-2)$-flats $<l_{i_1},l_{i_2},\cdots,l_{i_{k-4}},Q>$ with $1\leq i_1< i_2<\cdots< i_{k-4}\leq q-1$,
\item the $\binom{q-1}{k-4}$ many $(k-2)$-flats $<l_{i_1},l_{i_2},\cdots,l_{i_{k-4}},P_q>$ with $1\leq i_1< i_2<\cdots< i_{k-4}\leq q-1$,
\item the $\binom{q-1}{k-3}$ many $(k-2)$-flats $<l_{i_1},l_{i_2},\cdots,l_{i_{k-3}}>$ with $1\leq i_1< i_2<\cdots< i_{k-3}\leq q-1$.
\end{enumerate}
Then $a_{(k-1)q-2}=\binom{q-1}{k-3}+2\binom{q-1}{k-4}+\binom{q-1}{k-5}$.
\end{proof}

Considering the $(k-2)$-flats with multiplicity $((k-1-j)q-1)$ in $\Sigma$ as $j$-points in $\Sigma^{*}$ for $j=0,1,2,\cdots,k-2$, one has the following $q^{k-3}$-divisible code $\mathcal {C}_2^{*}$ as a projective dual of $\mathcal {C}_2$.

\begin{lemma}
There exists a $q^{k-3}$-divisible $[3q^{k-1}-2q^{k-2}+1+(k-5)\theta_{k-1}, k,(k-2)q^{k-1}-5q^{k-2}+2q^{k-3}]_q$ code $\mathcal {C}_2^{*}$.
\end{lemma}

\begin{lemma}\label{SK2}
The multiset $\mathcal {M}_{\mathcal {C}_2^{*}}$ contains $(q-1)$ disjoint lines.
\end{lemma}

\begin{proof}
Note that the points with multiplicity 0 for $\mathcal {C}_2^{*}$ are the $(k-2)$-flats with multiplicity $((k-1)q-1)$ for $\mathcal {C}_2$.  Since $l_0^{*}$ is contained in $H_{i_1i_2\cdots i_{k-3}}$,
$<H_{i_1i_2\cdots i_{k-4}},P_q>$,
$<H_{i_1i_2\cdots i_{k-5}},P_q,Q,l_0>$
and $<H_{i_1i_2\cdots i_{k-4}},Q>$ in $\Sigma$, the $(k-3)$-flat $l_0^{*}$ contains exactly $\binom{q-1}{k-3}+2\binom{q-1}{k-4}+\binom{q-1}{k-5}$ many points with multiplicity 0 in $\Sigma^{*}$.
Hence the number of points with multiplicity $i\geq 1$ on $l_0^{*}$ is $\theta_{k-3}-\binom{q-1}{k-3}-2\binom{q-1}{k-4}-\binom{q-1}{k-5}\geq q-1$.
Recall that the $(k-3)$-flat $l_0^{*}$ is contained in the $(k-2)$-flats $P_0^{*}$ and $Q_1^{*},\cdots, Q_q^{*}$ in $\Sigma^{*}$. One can take $q-1$ skew lines in the $(k-2)$-flat $Q_1^{*}$ containing no point with multiplicity 0 in $\Sigma^*$.
\end{proof}

\noindent \textbf{Proof of Theorem 1}

Starting with the code $\mathcal {C}_1^*$, by Lemma \ref{SK1} and applying Lemma \ref{t=1} repeatedly, we get the following.

\textbf{Fact1.}
There exists $[2q^{k-1}-q^{k-2}+1+(k-5)\theta_{k-1}-s(q+1),k,(k-3)q^{k-1}-3q^{k-2}+q^{k-3}-sq]_q$
code for every $k\geq 6$ and $1\leq s\leq q-1$.

By puncturing these divisible codes, we obtain the following.

\textbf{Fact2.} There exists $[g_q(k,d),k,d]_q$ code for $(k-3)q^{k-1}-3q^{k-2}+q^{k-3}-q^2+q\leq d\leq (k-3)q^{k-1}-3q^{k-2}+q^{k-3}$ for every $k\geq 6$ and $q\geq k-2$.

Combining Fact2 with Theorem \ref{k51} for $k=5$ yields Theorem 1.$\hfill\square$

\noindent \textbf{Proof of Theorem 2}

Starting with the code $\mathcal {C}_2^*$, by Lemma \ref{SK2} and applying Lemma \ref{t=1} repeatedly,  we get the following.

\textbf{Fact3.}
 There exists $[3q^{k-1}-2q^{k-2}+1+(k-5)\theta_{k-1}-s(q+1),k,(k-2)q^{k-1}-5q^{k-2}+2q^{k-3}-sq]_q$
code for every $k\geq 6$ and $1\leq s\leq q-1$.

By puncturing these divisible codes, we obtain the following.

\textbf{Fact4.} There exists $[g_q(k,d),k,d]_q$ code for $(k-2)q^{k-1}-5q^{k-2}+2q^{k-3}-q^2+q\leq d\leq (k-2)q^{k-1}-5q^{k-2}+2q^{k-3}$ for every $k\geq 6$ and $q\geq max\{5,k-2\}$.

Combining Fact4 with Theorem \ref{k52} for $k=5$ yields Theorem 2. $\hfill\square$

\begin{example}
Take $q=4$ and $k=6$. Then we can get Griesmer codes with parameters in Table 1 by Theorem \ref{the1}.
\begin{table}[H]
\setlength{\belowcaptionskip}{0.2cm}
\caption{}\label{table 1}
    \centering
    \begin{tabular}{|l|l|}
    \hline
        $n=g_4(6,d)$ & $d$ \\ \hline
        3158 & 2368 \\ \hline
        3157 & 2367 \\ \hline
        3156 & 2366 \\ \hline
        3155 & 2365 \\ \hline
        3153 & 2364 \\ \hline
        3152 & 2363 \\ \hline
        3151 & 2362 \\ \hline
        3150 & 2361 \\ \hline
        3148 & 2360 \\ \hline
        3147 & 2359 \\ \hline
        3146 & 2358 \\ \hline
        3145 & 2357 \\ \hline
        3143 & 2356 \\ \hline
    \end{tabular}
\end{table}
\end{example}

\begin{example}
 Take $q=5$ and $k=6$. Then we can get Griesmer codes with parameters in Table 2 by Theorem \ref{the2}.
\begin{table}[H]
\setlength{\belowcaptionskip}{0.2cm}
\caption{}\label{table 2}
    \centering
    \begin{tabular}{|l|l|}
    \hline
        $n=g_5(6,d)$ & $d$ \\ \hline
        12032 & 9625 \\ \hline
        12031 & 9624 \\ \hline
        12030 & 9623 \\ \hline
        12029 & 9622 \\ \hline
        12028 & 9621 \\ \hline
        12026 & 9620 \\ \hline
        12025 & 9619 \\ \hline
        12024 & 9618 \\ \hline
        12023 & 9617 \\ \hline
        12022 & 9616 \\ \hline
        12020 & 9615 \\ \hline
        12019 & 9614 \\ \hline
        12018 & 9613 \\ \hline
        12017 & 9612 \\ \hline
        12016 & 9611 \\ \hline
        12014 & 9610 \\ \hline
        12013 & 9609 \\ \hline
        12012 & 9608 \\ \hline
        12011 & 9607 \\ \hline
        12010 & 9606 \\ \hline
        12008 & 9605 \\ \hline
    \end{tabular}
\end{table}
\end{example}

\end{document}